\documentclass[aps,prd,twocolumn]{revtex4}
\usepackage{amsmath,amssymb,amsfonts}
\usepackage{graphicx,epsfig,graphics,rotate}
\usepackage{dcolumn}
\usepackage{bm}
\usepackage{epsfig}

\newcommand{\lie}{{\mathcal{L}}}
\def\dd{{\rm d}}

\begin{document}
\title{Quantization of perturbations during inflation in the $1+3$ covariant formalism}

\author{Cyril Pitrou}
 \email{pitrou@iap.fr}
 \affiliation{Institut d'Astrophysique de Paris,
              UMR-7095 du CNRS, Universit\'e Pierre et Marie
              Curie,
              98 bis bd Arago, 75014 Paris (France)}

\author{Jean-Philippe Uzan}
 \email{uzan@iap.fr}
 \affiliation{Institut d'Astrophysique de Paris,
              UMR-7095 du CNRS, Universit\'e Pierre et Marie
              Curie,
              98 bis bd Arago, 75014 Paris (France)}

\date{\today}
\begin{abstract}
This note derives the analogue of the Mukhanov-Sasaki variables both for scalar and tensor perturbations in the 1+3 covariant formalism. The possibility of generalizing them to non-flat Friedmann-Lema\^{\i}tre universes is discussed.
\end{abstract}

\maketitle

The theory of cosmological perturbations is a cornerstone of the modern cosmological model. It
contains two distinct features. First, it describes the growth of density perturbations from an
initial state in an expanding universe filled with matter and radiation, taking into account the
evolution from the super-Hubble to the sub-Hubble regime. Second, it finds the origin of the
initial power spectrum of density perturbations in the amplification of vacuum quantum fluctuations
of a scalar field during inflation.

Two formalisms are used to describe the evolution of perturbations. The first relies on the
parametrization of the most general spacetime close to a Friedmann-Lema\^{\i}tre (FL) universe and
on the construction of gauge invariant variables~\cite{bardeen81}. The second is based on a
general $1+3$ decomposition of the Einstein equation~\cite{ellis71}. The philosophies and advantages
of these two formalisms are different.

The Bardeen formalism is restricted to perturbations around a FL universe (with metric $\bar
g_{\mu\nu}$) and expands the spacetime metric as $g_{\mu\nu} = \bar g_{\mu\nu}+\gamma_{\mu\nu} $. The
metric perturbation is then decomposed as
\begin{eqnarray}
\gamma_{\mu\nu}\dd x^\mu\dd x^\nu = 2a^2(\eta)\left[-A\dd\eta^2 + \left(D_iB+\bar B_i\right)\dd
x^i\dd\eta \right.\nonumber\\
\left.+ \left( C\gamma_{ij}+ D_iD_jE+D_{(i}\bar E_{j)}+\bar E_{ij}\right)\dd x^i\dd x^j\right]
\label{decmet}
\end{eqnarray}
and the 10 degrees of freedom are decomposed in 4 scalars ($A,C,B,E$), 4 vectors ($\bar E_i,
\bar B_i$ with $D_i\bar B^i=0,\ldots$) and 2 tensors ($\bar E_{ij}$ with $\bar E_{i}^i=D_i\bar
E^{ij}=0$). During inflation, it was shown that the Mukhanov-Sasaki (MS) variable $v$, a gauge invariant
variable that mixes matter and metric perturbations, must be quantized~\cite{mukhanov,sasaki,mbf}. This approach is
thus completely predictive (initial conditions and perturbations evolution) for an almost flat FL
spacetime. It is not straightforward to extend it to less symmetrical inflationary
models~\cite{inflK,anisoinfl}, for which the quantization procedure has not been investigated (see
however Ref.~\cite{Kneg} for non flat FL universes and Ref.~\cite{bardeenaniso} for non-FL
spacetimes).

The $1+3$ covariant description assumes the existence of a preferred congruence of worldlines
representing the average motion of matter. The central object is the 4-velocity $u^a$ of these
worldlines together with the kinematical quantities arising from the decomposition
\begin{equation}\label{dec31}
 \nabla_a u_b = -u_a\dot u_b + \frac{1}{3}\Theta h_{ab} + \sigma_{ab}
 +\omega_{ab}
\end{equation}
where $h_{ab}=g_{ab}+u_a u_b$. $\Theta=\nabla_a u^a$ is the
expansion rate,
$\sigma_{ab}$ the shear (symmetric trace-free with $\sigma_{ab}u^a=0$),
$\omega_{ab}$ the vorticity (antisymmetric with $\omega_{ab}u^a=0$) and $\dot
u_a\equiv u^b\nabla_b u_a$). A fully orthogonally projected covariant derivative $D_a$, referred to as a spatial gradient, is defined and a complete set of evolution equations can be obtained for
these quantities (together with the electric and magnetic parts of the Weyl tensor, $E_{ab}$
and $H_{ab}$ and the matter variables) without needing to specify the spacetime geometry (see
e.g. Ref.~\cite{elst} for details). The formalism can then be used to study the evolution of
perturbations in various spacetimes (see e.g. Ref.~\cite{dunsby1} for Bianchi universes) and
beyond the linear order in the case of almost FL universes~\cite{vernizzi1,vernizzi2}). In this
case, the perturbation variables have a clear interpretation and have been related to the Bardeen
variables both for fluids~\cite{covlien} and scalar fields~\cite{covinflation}. On the other hand,
the scalar-vector-tensor decomposition is not straightforward~\cite{gw} and the analogue of the
Mukhanov-Sasaki variable $v$ has not been derived so that it is difficult to argue which quantity must be quantized in this formalism.

The goal of this article is to identify this variable in the $1+3$ covariant formalism. In
\S~\ref{sec1}, we recall the construction of the Mukhanov-Sasaki variable. We then identify in \S~\ref{sec2} its
counterpart in the $1+3$ covariant formalism for an almost FL spacetime. \S~\ref{sec3} adresses the
question of gravitational waves. Finally we conclude on the use and possible
extensions of these variables in \S~\ref{sec4}.

\section{Quantization of the Mukhanov-Sasaki variable}\label{sec1}

Focusing on scalar modes of the decomposition~(\ref{decmet}), one defines the two gauge invariant
potentials, $\Phi=A+\mathcal{H}(B-E')+(B-E')'$ and $\Psi=-C-\mathcal{H}(B-E')$ where
$\mathcal{H}=a'/a$. For a universe filled by a single scalar field $\varphi$, one can define the
gauge invariant field fluctuation $Q=\delta\varphi-\varphi'C/\mathcal{H}$. The curvature
perturbation in the comoving gauge is then given by
\begin{equation}
 \mathcal{R}=-C+\mathcal{H}\frac{\delta\varphi}{\varphi'} = \frac{\mathcal{H}}{\varphi'}Q,
\end{equation}
when assuming a background FL universe with flat spatial sections. Introducing the two variables
\begin{equation}
 u \equiv \frac{2 a \Phi}{\kappa\varphi'}\,,
 \qquad
 \theta \equiv \frac{1}{z} \equiv \frac{\mathcal{H}}{a\varphi'}=\frac{H}{a\dot\varphi},
\end{equation}
with $\kappa=8\pi G$, $\mathcal{R}$ takes the simple form $\mathcal{R}=\theta u' -
\theta' u$, where use has been made of the background equation $\theta'= \left(-a\theta\varphi'' + \mathcal{H}'-\mathcal{H}^{2}\right) /a\varphi' =
-\theta\varphi''/\varphi' - \kappa\varphi'/2a$.   If we
define
\begin{equation}
 v\equiv z\mathcal{R}=aQ
\end{equation}
it reduces to the simple relation $zv=(uz)'$. Thus, $\mathcal{R}'$ takes the simple form
$\mathcal{R}'=\theta u'' - \theta'' u$. Additionnaly, the perturbed Einstein equations imply that
$\mathcal{R}'=\theta \Delta u$. Provided $\varphi'\neq 0$, i.e. we are not in a strictly de Sitter
phase, this leads to
\begin{equation}
 u'' - \left(\Delta + \frac{\theta''}{\theta}\right) u = 0
\end{equation}
which can also be written as
\begin{equation}\label{e7}
 z^2 \mathcal{R}'=\Delta\left(\int \mathcal{R} z^2 \dd\eta\right).
\end{equation}
This form is strictly equivalent to $\mathcal{R}''- \Delta \mathcal{R} + 2 (z'/z) \mathcal{R}'=0$,
that is to
\begin{equation}\label{eqofv}
 v'' - \left(\Delta+\frac{z''}{z}\right) v = 0.
\end{equation}
This is the equation of a harmonic oscillator with time varying mass $m^2=z''/z$. To show that $v$
is indeed the canonical variable to be quantized (contrary to $u$ which satisfies a similar equation with $z$ replaced by $\theta$), one needs to expand the action of the scalar field coupled to gravity up to second
order~\cite{mbf}. The standard procedure is then to promote $v$ to the status of an operator and
set the initial conditions by requiring that it is in its vacuum state on sub-Hubble scales (in
Fourier space, that is when $k\eta\rightarrow-\infty$). At the end of inflation, all
cosmologically relevant scales are super-Hubble ($k\eta\ll1$) and the conservation of
$\mathcal{R}$ on these scales is used to propagate the large scale perturbations generated
during inflation to the post-inflationary eras.

\section{Quantization in the 1+3 formalism}\label{sec2}

\subsection{Generalities}

Let us first stress an important point concerning time derivative. In the coordinate based
approach, the dot refers to a derivative with respect to the cosmic time. In the covariant
formalism, a natural time derivative is introduced as $u^a\nabla_a$ which is a derivative along
the worldline of the observer. For an almost FL spacetime, it is clear that for any order 1 scalar quantity $X$, $u^a\nabla_a X=\partial_tX$ since $u^a$ has to be evaluated at the background level. It is
indeed not the case anymore for vector or tensor quantities, or at second order in the
perturbations. Another time derivative can be constructed from the Lie derivative along $u^a$,
$\mathcal{L}_u$. This derivative exactly matches the derivative with respect to cosmic time for
any type of perturbations~\cite{vern3}. For instance, $\mathcal{L}_u X_a =u^{b} \nabla_{b}
X_a + X_{b} \nabla_a u^{b} $, and it is easily checked that the second term exactly
compensates the term arising from the divergence of $u^a$. We recall that the Lie derivative
satisfies
\begin{equation}\label{idlie}
 D_a ( \dot{X}) = \lie_u (D_a X) - \dot u_a \dot{X}.
\end{equation}

In the $1+3$ covariant formalism, the main equations for a spacetime with vanishing vorticity
($\omega_{ab}=0$ in the decomposition~(\ref{dec31})) are the Raychaudhuri equation
\begin{equation}\label{raychau}
 \dot\Theta+\frac{1}{3}\Theta^2=D_a\dot u^a-\frac{\kappa}{2}(\rho+3P)-2\sigma^2,
\end{equation}
the Gauss-Codacci equation 
\begin{equation}\label{gauss}
 {\cal K}=2\left(-\frac{\Theta^2}{3}+\kappa\rho+\sigma^2\right),
\end{equation}
together with the conservation equations
\begin{equation}\label{conseq}
 \dot\rho+\Theta(\rho+P)=0,\quad
 D_aP=-(\rho+P)\dot u_a
\end{equation}
where $2\sigma^2=\sigma_{ab}\sigma^{ab}$. For a scalar field $\rho=\psi^2/2+V$ and $P=\psi^2/2-V$
with $\psi=\dot\phi$. We define the scale factor $S$ by the relation $\dot S/S=H\equiv\Theta/3$.
The previous equations imply that
\begin{equation}\label{Hdot}
 \dot H = -\frac{\kappa}{2}\psi^2-\sigma^2+\frac{1}{6}\mathcal{K}+D_a\dot u^a
\end{equation}
and the Gauss equation leads to
\begin{equation}\label{Kdot}
 \dot{\mathcal{K}}= -\frac{2}{3}\Theta(\mathcal{K}+2D_a\dot u^a)+2(\sigma^2)^. +2\Theta\sigma^2
\end{equation}
and
\begin{equation}\label{DaK}
 D_a{\mathcal{K}}= -\frac{4}{3}\Theta D_a\Theta+2\kappa\psi D_a\psi+2D_a(\sigma^2).
\end{equation}

Let us now introduce the central quantity in our discussion
\begin{equation}\label{defva}
 \mathcal{R}_{a} \equiv  \frac{1}{3} \left[\int_{\mathcal{L}} \left(D_a \Theta \right)\dd\tau - D_a
\left( \int_{\mathcal{L}} \Theta  \dd\tau \right) \right] \equiv \frac{v_a}{z}
\end{equation}
where $\tau$ is the proper time along the fluid flow lines. It follows from the
identity~(\ref{idlie}) that 
\begin{equation} \label{Radot}
\lie_u\mathcal{R}_a=-H\dot u_a=H(D_a\psi)/\psi,
\end{equation} 
which is a useful relation in order to close the equations in $\mathcal{R}_a$.
The conservation equation implies that $\lie_u(\psi D_a\psi)=-\psi D_a(\Theta\psi)$ which, combined with
Eq.~(\ref{DaK}), gives
\begin{equation}\label{e8}
 \frac{1}{4}S^2D_a(\mathcal{K}-2\sigma^2)=U_a+ \frac{2\Theta}{3\kappa\psi^2}\left[
 \lie_u U_a+\frac{1}{3}\Theta U_a \right]
\end{equation}
with $U_a=\kappa S^2\psi D_a\psi/2$. Now, introducing
\begin{equation}\label{defVa}
 V_a\equiv z S^2\left[
 \frac{1}{4}D_a(\mathcal{K}-2\sigma^2)+
 (\sigma^2-\frac{1}{6}\mathcal{K}-D_b\dot u^b)\frac{D_a\psi}{\psi}
 \right]
\end{equation}
and developping $\lie_u(z U_a)$ with the help of Eqs.~(\ref{e8}) and~(\ref{Hdot}), it can be
shown that
\begin{equation}\label{RaofVa}
 \lie_u\mathcal{R}_a =\frac{1}{S z^2}\int\frac{z V_a}{S}\dd\tau.
\end{equation}
Whatever the values taken by the shear and the spatial curvature, this
intermediate result is exact, that is
valid for any order of perturbation. Note its similarity with Eq.~(\ref{e7}).

\subsection{Flat FL spacetimes}

Let us now focus on homogeneous spacetimes with flat spatial sections; this includes the FL
spacetimes for which the shear vanishes and Bianchi~I spacetimes. Homegeneity implies that the
spatial gradient of any scalar function vanishes ($D_af=0$, $\forall f$) and in particular $D_aP=0$ so
that $\dot u_a=0$. Flatness implies that $\mathcal{K}=0$ and that the 3D-Ricci tensor vanishes,
$^{(3)}R_{ab}=0$, which leads to the simplified equation for the shear
\begin{equation}\label{shearevo}
 \dot\sigma_{ab}+\Theta\sigma_{ab}=0\, \Rightarrow \,\left(\sigma^2\right)^{.} + \Theta \sigma^2=0.
\end{equation}
This means that the only non-vanishing quantities are $\Theta$, $\phi$, $\psi$
and $\sigma$. At
first order in the perturbations, we have to consider the gradients of these quantities but also
terms like $\dot u_a$ and $\mathcal{K}$.\\
FL spacetimes are also isotropic. This implies that $\sigma=0$ and $S=a$ for the
background. Consequently we can discard gradients of $\sigma^2$ which are second order. Thus, the only remaining
term in the expression~(\ref{defVa}) of $V_a$ is $\frac{1}{4}z S^2 D_a \mathcal{K} $. 

In order to get a closed equation for $\mathcal{R}_a$, we need to express $V_a$
in terms of $\mathcal{R}_a$ in Eq.~(\ref{RaofVa}). Taking the spatial gradient of
Eq.~(\ref{Kdot}) and using Eq.~(\ref{idlie}) for handling the time derivative, we obtain the first order relation
\begin{equation}
\lie_u\left(\frac{S^2 D_a \mathcal{K} }{4}\right) = \frac{S^2 H}{\psi} D_a\left( D_b
  D^b \psi\right) = S^2 H D_b D^b \left( \frac{D_a \psi}{\psi}\right),\nonumber 
\end{equation}
where the last equality follows from the flat background assumption. Eq.~(\ref{Radot}) and the
commutation relation $S^2 D_b D^b \lie_u X_a = \lie_u \left(S^2 D_b D^b
  X_a\right)$, valid at first order, imply $\lie_u\left(\frac{S^2 D_a
    \mathcal{K} }{4}\right) = \lie_u \left( S^2 D_b D^b \mathcal{R}_a \right)$
which once integrated leads to

\begin{equation}\label{KatoRa}
\frac{S^2 D_a
    \mathcal{K} }{4} = S^2 D_b D^b \mathcal{R}_a \equiv \Delta \mathcal{R}_a
\end{equation}
up to a constant $F_a$ satisfying $\lie_u F_a =0$ which can be absorbed in the
integration initial boundary surface of Eq.~(\ref{defva}). Thus Eq.~(\ref{RaofVa}) reads 
\begin{equation}\label{Raclosed}
 \lie_u\mathcal{R}_a =\frac{1}{Sz^2}\int \frac{z^2}{S}\Delta \mathcal{R}_a  \dd\tau.
\end{equation}
At this stage, it is useful to introduce a vector field $w_{a} \equiv S
u_{a}$, and the conformal proper time $\hat \tau$ defined by $S \dd\hat \tau = \dd
\tau$. It is easily seen that for a spatial vector
(i.e. $u^{a}X_{a}=0$) $\lie_w X_a = S \lie_u X_a $. The Lie derivative along
$w_{a}$ matches at first order the derivative with respect to the conformal
time $\eta$, as the Lie derivative along $u_{a}$ was matching the derivative
with respect to the cosmic time. For scalars we thus use the notation $X' \equiv \lie_w X$. With this definition, Eq.~(\ref{Raclosed}) can be recast as

\begin{equation}
z^2 \lie_w\mathcal{R}_a =\int z^2\Delta \mathcal{R}_a \dd \hat\tau,
\end{equation}
which is similar to Eq.~(\ref{e7}). By the same token, we deduce
that $v_a$ defined in Eq.~(\ref{defva}) satisfies Eq.~(\ref{eqofv}).  
It can be checked that its spatial components are linked to the
MS variable at first order in perturbations by $v_i= \partial_i
v$, and consequently the initial conditions obtained from the quantization of $v$
can be used to set the initial conditions for $v_a$ and then $\mathcal{R}_a$. $v_a$
 is thus the analogue in the $1+3$ formalism of the MS variable $v$ in the
 Bardeen formalism and it satisfies
\begin{equation}
\lie_w^2 \left(v_{a} \right)-\left(\Delta + \frac{z''}{z}\right) v_{a}=0\,.
\end{equation}

\section{Gravitational waves} \label{sec3}

It can be shown that the magnetic part of the Weyl tensor $H_{ab}$ is a good variable to describe the
gravitational waves~\cite{gw,OPDUC}, and it satisfies at first order for flat FL spacetimes

\begin{equation}
\lie_w^2 H_{ab} + 2 \lie_w \left( \mathcal{H} H_{ab} \right) - \Delta H_{ab}=0\,.
\end{equation}
$\mathcal{E}_{ab}$ defined by $\lie_w \mathcal{E}_{ab} = H_{ab}$ satisfies 
\begin{equation} \label{dynEab}
\lie_w^2 \mathcal{E}_{ab} + 2 \mathcal{H} \lie_w \left( \mathcal{E}_{ab} \right) - \Delta \mathcal{E}_{ab}=0
\end{equation}
where the integration constant is set to $0$ as for the scalar case. 
This is exactly the equation satisfied by the gravitational waves
$\bar E_{ij}$. Indeed, these variables are linked at first order by $\mathcal{E}_{ab}=
{\epsilon^{cd}}_{<a}\partial_c \bar E_{b>d} $~\cite{OPDUC}, where ${\epsilon^{ab}}_{c}$ is a
 completely antisymmetric tensor normalized such that ${\epsilon^{12}}_{3}=1$. With $\mu_{ab} \equiv \frac{S}{\sqrt{8
    \pi G}} \mathcal{E}_{ab}$, Eq.~(\ref{dynEab}) leads at first order to
\begin{equation}
\lie_w^2 \left(\mu_{ab}\right)-\left(\Delta + \frac{S''}{S}\right) \mu_{ab}=0
\end{equation}  
which is the equation for an harmonic oscillator with a time varying mass
$S''/S$. Just as for scalar perturbations case, the quantization of the gravitational
waves in perturbation theory, can be used to set the initial conditions for $\mu_{ab}$ and thus for $\mathcal{E}_{ab}$ and $H_{ab}$.  

\section{Conclusions}\label{sec4}

We have identified the scalar and tensor variables that map to the Mukhanov-Sasaki variables when considering an
almost-FL universe with Euclidean spatial sections in the $1+3$ covariant formalism. Let us stress
that in Ref.~\cite{vernizzi1}, $\mathcal{R}_a \equiv -D_a \alpha$ (where $\alpha=
\frac{1}{3}\int_{\mathcal{L}} \Theta \dd\eta$ is the integrated volume expansion along $u_a$) was proposed. But
clearly, this maps at linear order to $\partial_a \left(\mathcal{R} - \frac{1}{3}\int \Delta
(V+E') \dd\eta \right)$, where $V$ is the scalar part of the perturbation of the
velocity. The additional term in our definition~(\ref{defva}) cancels this discrepancy as it can be seen from the constraint $
 \frac{2}{3}D_a \Theta = D_b \sigma^b_{\,\,a} $ which implies that at first order $D_a \Theta = \frac{1}{a} \partial_{a} \Delta V$. Alternatively, this can be seen
directly on the expression of $\Theta$ in terms of perturbation variables by
use of the ($0-i$) Einstein equation. The two variables agree at leading order
on super-Hubble scales. 

Two generalizations with less restrictive backgrounds can be considered:
flat but anisotropic spatial sections ($K\equiv \frac{S^2 \mathcal{K}}{6}
=0$, $\sigma \neq 0$), and isotropic but non-flat FL spacetimes ($K \neq 0$, $\sigma=0$). The cornerstone of the
derivation of \S~\ref{sec2} is the possibility of expressing $V_a$ only in terms of
$\mathcal{R}_a$ to get a closed equation from Eq.~(\ref{RaofVa}). \\
In the first case, two other terms in Eq.~(\ref{defVa}) contribute at first
order. $z S^2 \sigma^2 \frac{D_a \psi}{\psi}$ changes the definition of the
effective varying mass, but the term $\frac{z S^2}{2} D_a(\sigma^2) $ acts as
a source in the R.H.S of Eq.~(\ref{eqofv}). This term represents a coupling of
the gravitational waves ($\sigma_{ab}$ at first order) with the shear of the
background spacetime. As for $\mathcal{E}_{ab}$, Eq.~(\ref{dynEab}) will be
supplemented with an integral non trivial source term which couples the
background shear to the electric part of the Weyl tensor~\cite{OPDUC}. Thus
both equations are mixed with these new source terms, which are of the same
order of magnitude as the quantized variables. Note that this not surprising
since at second order in the perturbations around a FL spacetime the scalar and tensor degrees of freedom are coupled~\cite{maldacena03,OPDUC}. \\
In the second case ($K \neq 0$, $\sigma=0$),
\begin{equation}
V_a=z \left[\frac{S^2}{4}D_a
  \mathcal{K} - K \frac{D_a\psi}{\psi}\right] \equiv
\frac{z}{4}\tilde{C}_a  .
\end{equation} 
The spatial gradient of Eq.~(\ref{Kdot}) leads at first order to
\begin{equation} \label{evolutionCa}
\lie_u \tilde{C}_a = \frac{S^2 H}{\psi} D_a
\left(D_b D^b \psi\right) -3 K \lie_u \zeta_a  
\end{equation}
where $\zeta_a \equiv D_a \alpha + \frac{D_a \psi}{3 \psi} $ is a possible
nonlinear generalization of the curvature perturbation on uniform density hypersurfaces. From the
conservation equation~(\ref{conseq}) it can be shown~\cite{vernizzi2,vern3} that
$\zeta_a$ is conserved (in the sense  of the Lie derivative) on large scales for adiabatic perturbations, and thus $\tilde{C}_a$ is also conserved in the same
sense on super-Hubble scales for adiabatic perturbations~\cite{EBH}. Because of this
term, $\tilde{C}_a$ cannot be expressed solely in terms of $\mathcal{R}_a$ as it has been done
with Eq.~(\ref{KatoRa}). Indeed, it contains an additional term involving
$\zeta_a$~\cite{vernizzi1}.


\end{document}